\documentclass[twocolumn,PRL,showpacs,nofootinbib,amsmath,amssymb]{revtex4}

\usepackage{graphicx}
\usepackage{dcolumn}
\usepackage{bm}
\usepackage{hyperref}
\usepackage{color}
\usepackage{rotate}
\usepackage{fancyhdr} 
\usepackage{cancel}

\def\be{\begin{equation}}
\def\ee{\end{equation}}
\def\bea{\begin{eqnarray}}
\def\eea{\end{eqnarray}}

\def\VEV#1{\left\langle #1 \right\rangle}
\newcommand{\fnl}{{f_{\rm nl}}}

\definecolor{darkred}{RGB}{175,0,0}
\definecolor{darkblue}{RGB}{14,0,185}

\newcommand{\refeq}[1]{Eq.~(\ref{eq:#1})}

\newcommand{\bfx}{\mathbf{x}}
\newcommand{\bfr}{\mathbf{r}}
\newcommand{\bfk}{\mathbf{k}}
\newcommand{\bgfk}{\mathbf{K}}

\newcommand{\bfeps}{\mathbf{\epsilon}}

\begin{document}

\title{Antisymmetric galaxy cross-correlations as a cosmological probe}

\author{Liang Dai$^1$, Marc Kamionkowski$^1$, Ely D. Kovetz$^1$, Alvise
     Raccanelli$^1$, and Maresuke Shiraishi$^2$}
\affiliation{
     $^1$Department of Physics and Astronomy, Johns Hopkins
     University, 3400 N.\ Charles St., Baltimore, MD 21218, USA \\
     $^2$Kavli Institute for the Physics and Mathematics of
     the Universe (Kavli IPMU, WPI), UTIAS, The University of
     Tokyo, Chiba, 277-8583, Japan}
\date{\today}

\begin{abstract}
The auto-correlation between two members of a galaxy population
is symmetric under the interchange of the two
galaxies being correlated.  The {\it cross}-correlation between
two different types of galaxies, separated by a vector $\bfr$,
is not necessarily the same as that for a pair separated by
$-\bfr$.  Local anisotropies in the two-point
cross-correlation function may thus indicate a specific
direction which when mapped as a function of position
trace out a vector field.  This vector field can then be
decomposed into longitudinal and transverse components, and
those transverse components written as positive- and negative-helicity
components. A locally asymmetric cross-correlation of the
longitudinal type arises naturally in halo clustering, even with
Gaussian initial conditions, and could be enhanced with
local-type non-Gaussianity.  Early-Universe scenarios that
introduce a vector field may also  give rise to such effects.
These antisymmetric cross-correlations also provide a new
possibility to seek a preferred cosmic direction correlated with
the hemispherical power asymmetry in the cosmic microwave
background and to seek a preferred location associated with the
CMB cold spot.  New ways to seek cosmic parity breaking are also
possible.
\end{abstract} 


\maketitle

\section{Introduction} Considerable evidence has accrued for the
past decade that large-scale structure in the Universe grew via
gravitational infall from a nearly scale-invariant spectrum of
very nearly Gaussian primordial adiabatic density perturbations
\cite{Bennett:2012zja,Ade:2015xua}.
The nature of these perturbations is generally accounted for in terms of
single-field slow-roll (SFSR) inflation \cite{inflation}.
However, even if SFSR is the correct explanation, there might
be new physics beyond SFSR inflation, and there remains
the ever-present possibility that primordial perturbations are
due to something completely different.

In order to develop a clearer understanding of the new physics
responsible for primordial perturbations, we must be vigilant in
seeking new fossils from the early Universe in the form of
subtle correlations in the matter distribution.  This is
the motivation for much of the work on non-Gaussianity
\cite{Bartolo:2004if}.
This {\it Letter} will explore new observables that can be sought
with galaxy surveys. Such work is timely given the advent in
the forthcoming years of a new generation of surveys
\cite{Ivezic:2008fe,Maartens:2015mra,Green:2012mj,Laureijs:2011gra} 
that will map the distribution of
galaxies over vast volumes in the Universe.  

Recent work \cite{Jeong:2012df} presented a parametrization of
the most general autocorrelation function---which may depend on
the orientation of the two points being correlated as well as
their position in space---and showed that it could be
decomposed into scalar, vector, and tensor components.
Here we generalize that work to the most
general two-point {\it cross}-correlation function between two
different galaxy populations.  If primordial perturbations are Gaussian,
then the two-point correlation function is statistically isotropic and
homogeneous.  More generally, though, the two-point
correlation function may, at least in some small region
of space, be anisotropic.  That anisotropy may also vary
from one point in the Universe to another.
Ref.~\cite{Jeong:2012df} parametrized the most general local
departure from isotropy for an auto-correlation function.  Since
an auto-correlation function for two galaxies separated by a
vector $\bfr$ must be invariant under the inversion $\bfr
\to -\bfr$, the most general departure from statistical
isotropy is parametrized in terms of a {\it symmetric} tensor, or
more generally, in terms of the six degrees of freedom that parametrize
that tensor.

If, however, we consider the two-point {\it cross}-correlation
function between two distinct populations, then it is possible
that the correlation function for a galaxy pair separated by
$\bfr$ may differ from that with separation $-\bfr$.  There
is thus a possibility that the two-point correlation function
might ``point'' in a given direction, and this pointing is
described by a vector.  Our purpose here is to
point out that evidence of the imprint of a vector field can be
sought with cross-correlatoins in the galaxy distribution.
Our work differs from that of Refs.~\cite{Bonvin:2013ogt,McDonald:2009ud,Yoo:2012se}, who
also considered asymmetric galaxy cross-correlations; while they
considered only those that arise from projection
effects, we consider {\it bona fide} asymmetries in the
three-dimensional distribution. Since observations are done in 
redshift space, though, care must be taken to disentangle the 
3D asymmetries we consider below from the redshift-space 
asymmetries discussed in Refs.~\cite{Bonvin:2013ogt,McDonald:2009ud,Yoo:2012se}, and from 
the angular dependence induced by the 3D RSD operator~\cite{Raccanelli:3D}.

Below we describe the parametrization of the most general
two-point cross-correlation, beginning by reprising the
parameterization for the auto-correlation function.  We then
discuss the decomposition of these types of local departures
from statistical isotropy in terms of two transverse and one
longitudinal mode. Next, we write down the optimal quadratic
estimators to be constructed from galaxy surveys to detect these
departures from statistical isotropy. After that, we discuss several
possible physical mechanisms to generate these types of local
departures from statistical isotropy.  We begin this discussion with an
effect that arises in biased halo clustering, even with Gaussian
initial conditions, and another that arises in halo clustering if
there is local-type primordial non-Gaussianity. We then describe
several early-Universe scenarios for such correlations.
We point out that a search for these anisotropic
cross-correlations may be used to seek a preferred cosmic
direction correlated with the direction indicated by the
hemispherical power asymmetry \cite{Eriksen:2003db,Ade:2013nlj}
or to seek a preferred location associated with the CMB cold
spot \cite{Vielva}.  Finally, we mention the possibliity to
construct new probes of parity breaking.

\section{Parametrization of the cross-correlation}  We begin by
reprising the parametrization of
Ref.~\cite{Jeong:2012df} for the two-point auto-correlation
function.  Suppose we have a density field $\delta(\bfx)$ from
which we construct the Fourier components
$\delta(\bfk)$.
The most general two-point correlation function for those 
Fourier components can be parametrized as,
\begin{eqnarray}
     \VEV{ \delta(\bfk_1) \delta(\bfk_2)} &=& P(k_1)
     \delta_{\bfk_1, -\bfk_2}
     + \sum_{\bgfk} \sum_p
     f_p(k_1,k_2,\mu) \nonumber \\
     & & \times h_p^*(\bgfk)
     \epsilon_{ij}^p(\bgfk) k_1^i k_2^j
     \,\delta^D_{\bfk_{123}},
\label{eq:fossil}
\end{eqnarray}
where $\delta_{\bfk_1, -\bfk_2}$ is a Kronecker delta, and
$\delta^D_{\bfk_{123}}$ is a Kronecker delta that sets $\bfk_1 +
\bfk_2 +\bgfk=0$.  Here $p$ sums over the six possible
basis tensors for a symmetric tensor, and $\epsilon_{ij}^p(\bgfk)$ are
the six polarization tensors which can be written as a trace, a
longitudinal mode, two transverse-vector modes, and two
transverse traceless tensor modes.  The $h_p(\bgfk)$ are
Fourier amplitudes of the various types of perturbations, and
$f_p(k_1,k_2,\mu)$ parametrizes the dependence
on $\bfk_1$ and $\bfk_2$ (where $\mu$ is the cosine of the
angle between $\bfk_1$ and $\bfk_2$).  The
first term in \refeq{fossil} indicates the statistically
isotropic correlation with power spectrum $P(k)$.  It was argued
in Ref.~\cite{Jeong:2012df}, simply from symmetry
considerations (and in particular, the dependence of the
correlation on the azimuthal angle about the direction of $\bgfk$), 
that terms with the two scalar, the two vector, or the two tensor
polarizations can arise in inflation if the inflaton is coupled
to a new scalar, vector, or tensor field, respectively.

Our purpose here, though, is to consider the additional
possibility that arises when we cross-correlate two different
tracers of the matter distribution. The fractional density
perturbations of these two populations will have Fourier
amplitudes $\delta_1(\bfk_1)$ and $\delta_2(\bfk_2)$, respectively.  The
most general two-point correlations of these Fourier
coefficients will satisfy a relation like \refeq{fossil}, but
the sum on $p$ will now be extended from six to nine to account
for the three possibilities where $\epsilon^p_{ij}(\bgfk)$ is
antisymmetric.  The three new degrees of freedom in the
cross-correlation, not allowed with auto-correlations, can be
written most generally, for $\bfk_1 \neq -\bfk_2$, as,
\begin{equation}
     \VEV{ \delta_1 (\bfk_1) \delta_2
     (\bfk_2)} =  \sum_{\bgfk,p} f_p\left(
     \bfk_1,\bfk_2,\mu \right) h_p^*(\bgfk) \hat \epsilon_p \cdot \left(
     \bfk_1-\bfk_2\right) \delta^D_{\bfk_{123}}.
\label{eq:newfossils}
\end{equation}
Here, the sum on $p$ runs over the three polarizations
$p=L,x,y$, where $\hat \epsilon_L(\bgfk) = \hat K$ and
$\epsilon_{x,y}(\bgfk)$ are two other unit vectors orthogonal
to $\hat K$ and to each other.  Just from symmetry
considerations, we expect that correlations with $p = L$ could
arise if there is some longitudinal-vector field---or
equivalently, the scalar field from which it is
derived---coupled to whatever physics determines the galaxy
distribution.  Similarly, distortions with $p=x,y$ would require
that the galaxy distribution was determined, at least in part,
by some transverse-vector field.

\refeq{newfossils} describes, in Fourier
space, cross-correlations between two different populations, that
are antisymmetric in the exchange of the two populations.  In
configuration space, these correlations trace out a vector
field, as follows:  In any small volume of the Universe, the
cross-correlation could ``point'' in some given direction; i.e.,
there could be a mean offset between galaxies of type 1 (e.g.,
more massive galaxies) and galaxies of type 2 (e.g., less massive
galaxies).  This thus implies a preferred direction in that
volume.  The generality of the parametrization in
\refeq{newfossils} allows for the possibility that this
preferred direction is spatially dependent in such a way that
global statistical isotropy is preserved on sufficiently large scales.
It also allows for the possibility, though, for a preferred
direction across the entire observable Universe.  This latter
possibility clearly requires some exotic new physics, but the
imprint of a {\it local} preferred direction is not at all
exotic, as we will see.

\section{Estimators}  We now describe how to measure these
vectorial cross-correlations from a galaxy survey.
We will consider the longitudinal and transverse components
separately and begin with longitudinal distortions, supposing
that $f_L(k_1,k_2,\mu)$ is specified.
Suppose further that the physical model specifies that the
Fourier components $h_L(\bgfk)$ are selected from a Gaussian
distribution with power spectrum $P_L(K)=A_L P_L^f(K)$ of
amplitude $A_L$ and fiducial $K$ dependence $P_L^f(k)$.
Following the same steps as in Ref.~\cite{Jeong:2012df}, the
optimal estimator for the amplitude $A_L$ is given by
\begin{equation}
     \widehat{ A_L} = \sigma_L^2 \sum_{\bgfk} \frac{
     P_L^f(K)}{ 2 \left[ P_L^n(K) \right]^2} \left(V^{-1}
     \left| \widehat{h_L(\bgfk)} \right|^2 - P_L^n(K) \right),
\end{equation}
and this amplitude is determined with inverse variance,
\begin{equation}
     \sigma_L^{-2} =
     (1/2) \sum_{\bgfk}\left[P_L^f(K) \right]^2
     \left[P_L^n(K)\right]^{-2}.
\end{equation}
Here,
\begin{eqnarray}
    \widehat{h_L(\bgfk)} &=& P_L^n(K) \sum_{\bfk}
    \frac{f_L \left(k, |\bgfk-\bfk|,\mu \right) \hat{K} \cdot (\bgfk
    + 2\bfk)}{2 V P_1^{\rm tot}(k)
    P^{\rm tot}_2(|\bgfk-\bfk| )} \nonumber \\
    & & \ \ \ \ \times \delta_1(\bfk) \delta_2(\bgfk-\bfk),
\label{eqn:hLestimator}
\end{eqnarray}
is the estimator for the amplitude $h_L(\bgfk)$, where here
$\mu = \hat{k} \cdot (\bgfk-\bfk)$, and the noise
power spectrum for $\widehat{h_L(\bgfk)}$ is
\begin{equation}
     P_L^n(K) = \left[ \sum_{\bfk} \frac{ \left| f_L
     \left(k,|\bgfk-\bfk|, \mu \right) \hat{K} \cdot (
     \bgfk + 2\bfk) \right|^2}{2 V P^{\rm tot}_1(k) P^{\rm
     tot}_2\left(|\bgfk -\bfk| \right)}
     \right]^{-1}.
\end{equation}
In these expressions, $V$ is the volume of the survey, the sums
are over the wavenumbers $\bfk$ for which the density
perturbations are measured; $P_1(K)$ and $P_2(K)$ are the power
spectra for the two populations; and $P^{\rm tot}$ is the total
power spectrum, the sum of the signal and the shot noise.
If instead we consider distortions of the transverse-vector
type, then the optimal estimator for the amplitude $A_T$ for the
power spectrum for the transverse-vector field will be as above,
but with $\hat K$ replaced by $\hat\epsilon_i({\bgfk})$ and an
additional sum over the two transverse polarizations. 

\section{Sources of asymmetric galaxy cross-correlations}
\subsection{Biased halo clustering}  We now show that asymmetric
cross-correlations arise in the standard model for biased halo
clustering from Gaussian initial conditions.
Let $\delta(\bfx)$ be the
primordial fractional mass-density
perturbation, and suppose that this density field is populated by
two different types of halos.  These could be, for example,
high-mass halos (e.g., those that house giant ellipticals,
clusters, etc.) and low-mass halos (e.g., those that house
Milky-Way type galaxies, dwarf galaxies, etc.).  Let $n_1(\bfx)$ 
be the fractional perturbation in the number density, as a
function of position $\bfx$, of halos of type 1 and similarly
for $n_2(\bfx)$.  Since the abundance of halos in some region
is a nonlinear function of the local mass density, there are
nonlinear relations \cite{Fry:1992vr}, $n_1(\bfx) = b_1 \delta(\bfx)+ c_1
[\delta(\bfx)]^2+\cdots$ and $n_2(\bfx) = b_2 \delta(\bfx)+ c_2
[\delta(\bfx)]^2+\cdots$ between the halo and mass densities,
where $b_1$ and $b_2$ are the usual linear-bias parameters, and
$c_1$ and $c_2$ are nonlinear-bias parameters.  The nonlinear
relation between $n_i$ and $\delta(\bfx)$ implies that there
will be a three-point correlation,
\begin{eqnarray}
&\resizebox{.96\hsize}{!}{$\VEV{n_1(\bfk_1) n_2(\bfk_2) \delta(\bgfk)} = 2 P(K)
     \left[b_2 c_1 P(k_2) + b_1 c_2 P(k_1) \right]
     \delta^D_{\bfk_{123}}.$} &\nonumber \\ 
&&
\label{eq:biasedhaloclustering}     
\end{eqnarray}
If we antisymmetrize in $\bfk_1$ and $\bfk_2$, we find
an antisymmetric part, $\propto 2 P(K) (b_2c_1 -
b_1 c_2) \left[ P(k_1) - P(k_2)\right]$ to this three-point
function.  We now consider the squeezed
limit, $\bfk \equiv \bfk_1 \sim -\bfk_2$ and $k \simeq k_1,
k_2 \gg K$, appropriate if we are
considering the small-scale clustering of halos in the presence
of a low-pass-filtered density field.  The part of the
bispectrum antisymmetric in $\bfk_1$ and $\bfk_2$ is then
$2 (b_1 c_2-b_2 c_1) P(K)[dP(k)/dk] \bgfk \cdot \hat \bfk$. 
We thus see that in the presence of a long-wavelength
mode of $\delta(\bfx)$, the high-pass-filtered $n_1(\bfx)$
and $n_2(\bfx)$ have a Fourier-space cross-correlation precisely
of the form in \refeq{newfossils} with $f_L = [dP(k)/dk](b_1 c_2 -
b_2 c_1)(K/k)$. The longitudinal, rather than
transverse, mode is as expected in the presence of the
long-wavelength density perturbation.

Existing measurements \cite{Verde:2001sf,Gil-Marin:2014sta,Marin:2013bbb,McBride:2010zn,Marin:2010iv} of nonlinear-bias
parameters from galaxy bispectra
\cite{Fry:1994my,Matarrese:1997sk,Buchalter:1999vc}
have relied entirely on
measurement of auto-correlations.
Cross-correlations have also been used to infer bias parameters
\cite{Croft:1998wq,Martinez:1998be,Knobel:2012xb,Shen:2012rr},
but those works considered only the linear bias.
If a galaxy sample is broken up into two different populations,
the linear- and nonlinear-bias parameters for the two populations
can be measured with the bispectra for the two different populations.
The symmetric part of the cross-correlation between these two
populations then provides a measurement of the combination
$b_1 c_2 + b_2 c_1$, and the antisymmetric part of this
cross-correlation provides $b_2 c_1-b_1
c_2$.  The latter provides
some incremental improvement in statistical power, and it can
also complement other measurements of the bias parameters and be used to
check the validity of the biasing-model assumptions behind the
analyses. 
Indeed, from Eq.~(\ref{eq:biasedhaloclustering}) and what follows, 
we see that for a power-law $P(k)$, the ratio between the symmetric and asymmetric signals
is $\propto(b_2 c_1-b_1
c_2)K/2(b_1 c_2 + b_2 c_1)k$, which is generically small in 
the squeezed limit $k\gg K$. However, for populations with negative
non-linear bias on certain scales (see e.\ g.\ \cite{Pollack:2013alj}), it is certainly possible that the bias parameters contrive to 
yield $b_1 c_2 + b_2 c_1\approx0$, in which case the asymmetric information
becomes particularly beneficial.

\subsection{Primordial non-Gaussianity}  Asymmetry may arise even with
linear biasing if there is primordial non-Gaussianity of the
local type. In this case, the linear-bias parameters for the two
halo populations may be scale dependent,
$n_1(\bfk_1)=b_1(k_1)\delta(\bfk_1)$ and
$n_2(\bfk_2)=b_2(k_2)\delta(\bfk_2)$. Here $b_i(k) \approx b_i +
2(b_i-1)\delta_c \fnl \Omega_m3(aH/k)^2/2$ with the standard
scale-independent bias $b_i$ and some collapse threshold
$\delta_c$~\cite{Dalal:2007cu,Matarrese:2008nc,Schmidt:2010gw}. On
the other hand, the
matter-density perturbation is expected to develop a nonzero
bispectrum $B(k_1,k_2,K)$ due to nonlinear gravitational
growth \cite{Fry:1983cj,Goroff:1986ep,Kamionkowski:1998fv} (this
should not be confused with the primordial bispectrum). 
The three-point function $\VEV{n_1(\bfk_1)
n_2(\bfk_2) \delta(\bgfk)} \propto b_1(k_1) b_2(k_2)
B(k_1,k_2,K)$ is thus noninvariant if we exchange $\bfk_1$ and
$\bfk_2$, and we again recover an antisymmetric correlation $\propto \left[b_1(k_1) b_2(k_2) -
b_1(k_2) b_2(k_1) \right] B(k_1,k_2,K)/P(K)$, with
\bea
f_L = 3(b_1-b_2) \delta_c \fnl \Omega_m \frac{(aH)^2 K}{k_1^2 k_2^2} \frac{B(k_1,k_2,K)}{P(K)}.
\eea

\subsection{Primordial longitudinal vector field}  We now describe an
exotic scenario that could give rise to
an anisotropic two-point correlation.  Consider a curvaton-like
inflation model in which primordial density perturbations are
due not to fluctuations in the inflaton, but to fluctuations in
some spectator field.  Suppose further that this curvaton-like
field is complex and coupled to a new $U(1)$ gauge field.  There
will then be charge-density fluctuations that arise during
inflation, and if the $U(1)$ symmetry is broken after inflation,
those charge-density fluctuations may survive.  It is then
conceivable that dark matter may be like baryonic matter in our
Universe, in which there are light electron-like dark-matter
particles and heavier proton-like dark-matter particles of
opposite charge.  If so, then galaxies that form in regions with
a positive charge-density fluctuation may be different than
those that form in regions with negative charge-density
fluctuations.  The cross-correlations between these two
different types of galaxies---which for example may appear to us
as brighter or fainter galaxies---should then trace out the
large-scale electric-type field associated with the
early-Universe $U(1)$ gauge field (see Fig.~\ref{fig:transversevectorfield}).  
Such a longitudinal-type
cross-correlation could in principle be distinguished from that
due to halo clustering from the different dependences on $\bfk$.  Unlike
the halo-biasing case, in which small-scale directionality
indicated by anisotropic cross-correlations are correlated with
the long-wavelength density field, these anisotropic
correlations would have no cross-correlation with the
large-scale density field.

\subsection{Primordial transverse vector field}  
\begin{figure}[b!]
\includegraphics[width=0.9\linewidth]{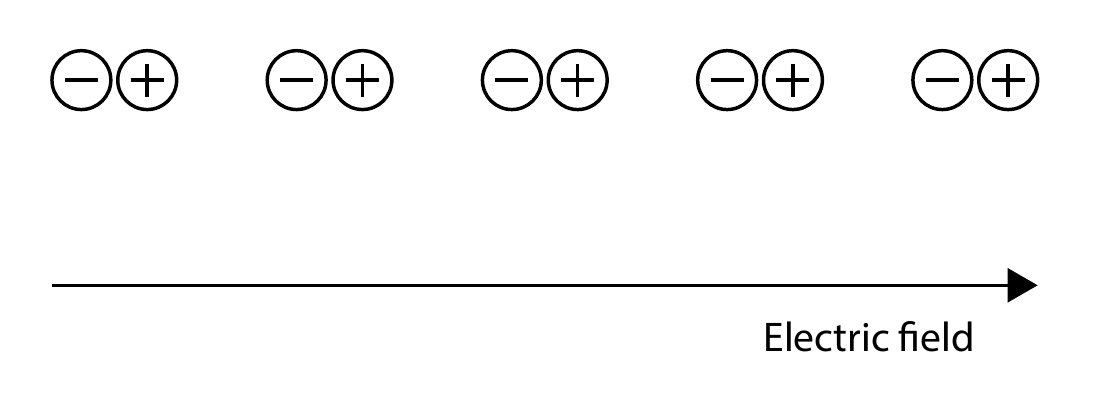}
\includegraphics[width=\linewidth]{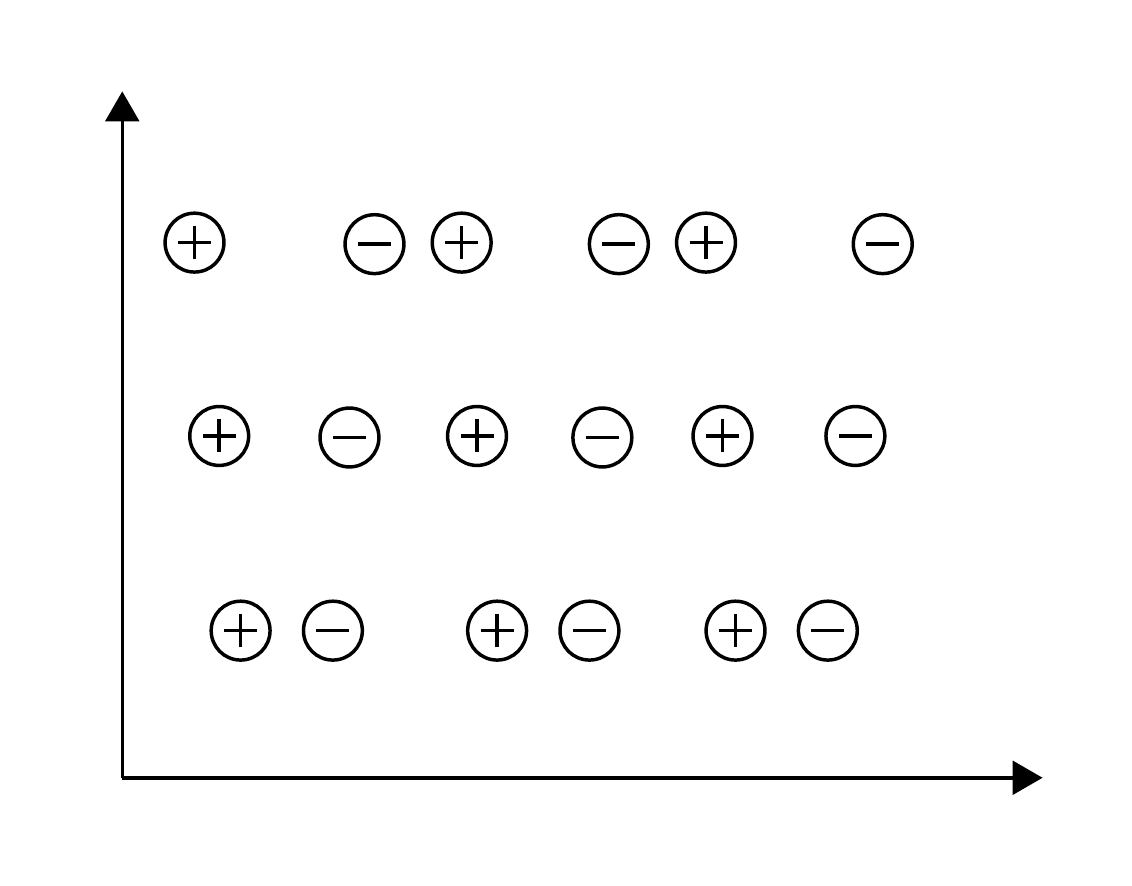}
\caption{We consider models with two components of dark matter with some dipole interaction. {\it Top:} An ``electric field" in the longitudinal direction leads to the arrangement in the pattern shown above. Then at the separation scale corresponding to the distance between ``-" and ``+", there is an asymmetric cross-correlation between the ``-" and ``+" populations, along this longitudinal direction. The estimators derived in this work are designed to detect such a transverse vector asymmetry. {\it Botom:} A different setup, for which the asymmetry has a sinusoidal modulation in the transverse direction, leading to a transverse asymmetric cross-correlation.}
\label{fig:transversevectorfield}
\end{figure}

What about
anisotropic cross-correlations of the transverse
variety?  As we have seen above, modulation of small-scale
galaxy-density fluctuations by a long-wavelength density mode
arise naturally, but those are of the longitudinal variety.  Any
small-scale anisotropic cross-correlation that traced out a
transverse-vector field would thus provide clear indication of
new physics.  There are indeed a number of inflationary
scenarios that involve new vector fields \cite{vectors} and
models of gravity in which the vector degrees of freedom in the
metric are brought to life \cite{Hellings:1973zz}, and it is
reasonable to surmise that one of these provide an imprint of
the type we consider.  To see how this may work, suppose the
inflaton (or curvaton) $\phi$ is
coupled to a vector field  $A^\mu$ through a term $\partial^{(\mu}
A^{\nu)} (\partial_\mu \phi) (\partial_\nu \phi)$, where the
parentheses in the superscripts indicate symmetrization
(antisymmetrization would make this term vanish).  Although
the coupling of a mode of wavenumber $\bgfk$ of the vector
field to two scalar-field modes of wavevectors $\bfk_1$ and
$\bfk_2$ would be symmetric under $\bfk_1 \leftrightarrow
\bfk_2$, one could again construct a model with two dark-matter
components, sourcing two different galaxy populations in the respective
dark matter halos made up from each component, in a way 
that makes the cross-correlation in one
of the transverse directions asymmetric (see illustration in Fig.~\ref{fig:transversevectorfield}).
These transverse-vector cross-correlations will
be straightforward to measure with a galaxy survey and any
non-null result would be extraordinarily interesting.  The
measurement will also be useful as a unique null test for
systematic effects.
 
\subsection{The CMB power asymmetry and cold spot} There is evidence
for a hemispherical power asymmetry in the cosmic microwave
background \cite{Eriksen:2003db,Ade:2013nlj} and it is important
to investigate whether this preferred direction shows up
anywhere else.  If the CMB power asymmetry is due to the
long-wavelength modulation of primordial perturbations
\cite{Prunet:2004zy,Gordon:2005ai,Gordon:2006ag,Erickcek:2008sm,Erickcek:2009at,Dai:2013kfa}, then
anisotropic cross-correlation of the type discussed above for
biased halo clustering should exist, but with just one
long-wavelength mode $h_L(\bgfk)$ of wavenumber $\bgfk\to 0$ in the
direction of the CMB power asymmetry.  In simple terms, this
Fourier amplitude $h_L(\bgfk)$ seeks whether a less massive
galaxy is more likely to be found on one side of a more massive
galaxy than on the other side.

Measurements from Planck \cite{Ade:2013nlj,Ade:2015hxq} also
confirm the existence of a cold spot in the CMB \cite{Vielva}.  An
asymmetric cross-correlation that traces out a hedgehog
configuration surrounding the cold spot could conceivably arise
either from a large void \cite{Inoue:2006rd} or from exotic physics \cite{Turok:1990gw} that may be
responsible for the cold spot.  A template for such a
configuration could be constructed from the estimators
$\widehat{h_L(\bgfk)}$ discussed above.

\subsection{Parity breaking?}  There are also novel probes of
cosmic parity breaking
that can be constructed.  Estimators for the two transverse
modes $h_p(\bgfk)$ for each wavevector $\bgfk$ can be
constructed analogously to Eq.~(\ref{eqn:hLestimator}) by
replacing the vector $\bgfk$ that appears in the dot product
with one of the two transverse polarization vectors.  Those two
linear polarization vectors $\hat \epsilon_x$ and $\hat
\epsilon_y$ can be replaced by circular polarizations $\hat
\epsilon_{\pm}(\bgfk) = 2^{-1/2}[ \epsilon_x(\bgfk) \pm
i \epsilon_y(\bgfk)]$.  Estimators for the Fourier amplitudes of
these helicity states are then given by
\begin{eqnarray}
    \widehat{h_\pm(\bgfk)} &=& P_L^n( K) \sum_{\bfk}
    \frac{f_L \left(k,  |\bgfk-\bfk|,\mu \right) \hat{\bfeps}_\pm 
    \cdot  (\bgfk + 2\bfk)}{2 V P_1^{\rm tot}(k)
    P^{\rm tot}_2(|\bgfk-\bfk| )} \nonumber \\
    & & \ \ \ \ \times \delta_1(\bfk) \delta_2(\bgfk-\bfk).
\label{eqn:hpmestimator}
\end{eqnarray}
It is then straightforward, following the
power-spectrum-amplitude estimators discussed above, to
construct estimators to check for an asymmetry between the power
in right and left circularly polarized modes.  Again,
while a null result would not be too surprising, a positive
result, if found, would be revolutionary.

\section{Conclusions} We have shown that cross-correlations
between different galaxy populations may ``point'' in a given
direction.  The vector field from this pointing
can be reconstructed from galaxy clustering, and it can be
decomposed into longitudinal and transverse components, and also
into components of positive and negative helicity.  We discussed
several physical mechanisms that may give rise to these
departures from local statistical isotropy.  These include
nonlinear halo biasing, an effect that should arise in
the standard model of halo clustering, and local-type non-Gaussian
primordial perturbations.  Other possiblities include couplings of the
primordial field responsible for primordial perturbations to a
new vector field, as well as a coupling to some field that might
account for the hemispherical power asymmetry or cold spot in the CMB.
The clustering effects from halo biasing constitute concrete
predictions of the standard halo-clustering model that can be
measured and used to test that standard model.  Although the
exotic possibilities we discussed are long shots, they will be
easily sought in forthcoming surveys and would, if discovered,
be quite remarkable.

{\it Acknowledgments.}  MK acknowledges the hospitality of the
Aspen Center for Physics, supported by NSF Grant No.\ 1066293.
This work was supported at JHU by NSF Grant No.\ 0244990, NASA
NNX15AB18G, the John Templeton Foundation, and the Simons
Foundation. MS was supported in part by a Grant-in-Aid for JSPS
Research under Grant No.~27-10917, and in part by World Premier
International Research Center Initiative (WPI Initiative), MEXT,
Japan.

\end{document}